\documentstyle[aps]{revtex}
\begin{document}
\begin{center}
{\bf Theory, Simulation and Nanotechnological Applications of
Adsorption on a Surface with Defects.}

Yu. E. Lozovik\footnote{e-mail: lozovik@isan.troitsk.ru,
tel: (07095) 3340881, fax: (07095) 3340886}, A. M. Popov

{\it Institute of Spectroscopy, Russian Academy of  Sciences,  142092
Troitsk, Moscow Region, Russia;}
\end{center}

\begin{abstract}
Theory of adsorption on a surface with nanolocal defects is
proposed. Two efficacy parameters of surface modification for
nanotechnological purposes are introduced, where the modification
is a creation of nanolocal artificial defects. The first
parameter corresponds to applications where it is necessary to
increase the concentration of certain particles on the modified
surface. And the second one corresponds to the pattern transfer
with the help of particle self-organization on the modified
surface. The analytical expressions for both parameters are
derived with the help of the thermodynamic and the kinetic
approaches for two cases: jump diffusion and free motion of
adsorbed particles over the surface. The possibility of selective
adsorption of molecules is shown with the help of simulation of
the adsorption of acetylene and benzene molecules in the pits on
the graphite surface. The process of particle adsorption from the
surface into the pit is theoretically studied by molecular
dynamic technique. Some possible nanotechnological applications
of adsorption on the surface with artificial defects are
considered: fabrication of sensors for trace molecule detection,
separation of isomers, and pattern transfer.
\end{abstract}


\vspace{0.5cm}

\section{Introduction.}

The progress in the development of nanotechnology gives rise to
the possibility to modify the surface at the nanometer scale.
Various methods of nanometer-scale defects creation using
scanning tunneling microscope (STM) and atomic force microscope
(AFM) were elaborated. In particular, nanometer-scale defects
can be produced by electric field between the STM needle tip and
the surface \cite{1a} -- \cite{1c}, with the help of nanolocal
chemical reactions induced by the STM needle tip \cite{2a} --
\cite{2c}, and by direct touching of the STM \cite{3a,3b} or AFM
\cite{3c} tip. Nanometer-scale defects may influence on
adsorption of particles (atoms, molecules and clusters)
\cite{4}. Surface nanostructures have been fabricated with the
help of adsorption on the surface modified by scanning probe
anodization \cite{4a}. We believe that the further advance in
nanotechnology can be achieved when techniques of nanostructures
fabrication with the help of adsorption will combine with
preliminary nanolocal surface modification. Therefore the
theoretical investigations of the nanolocal processes during
adsorption on heterogeneous surface is an actual task at present
time.

Here we present the theory of adsorption on the surface with
nanolocal defects, with sizes of the order of the size of only
one adsorption site (see also \cite{5,6}). Two efficacy
parameters of surface modification by creation of artificial
defects are introduced. The first efficacy parameter $\gamma$
can be useful for such applications where it is important to
increase the total concentration of particles on the surface. For
example, such applications may be fabrication of sensors for
trace molecule detection and separation of isomers. The second
efficacy parameter $\Gamma$ can be useful for applications
where high filling of adsorption sites on defects $\nu_d \approx
1$ is necessary simultaneously with the absence of adsorbed
particles on the surface free of defects. One of such
applications, for example, may be pattern transfer through the
use of particle self-organization on the modified surface. Two
limiting cases of adsorption are considered: the case of low
temperature when adsorbed particles are localized and the case of
high temperature when adsorbed particles freely move over the
surface. Analytical expressions for parameters $\gamma$ and
$\Gamma$ are derived for both cases of localized and freely
moving particles using equilibrium thermodynamics and the
kinetics principles.

The energy of molecular adsorption in pits with various sizes
and shapes is investigated. It is shown that this energy
is determined by the size of the pit. Therefore molecular
adsorption on defects may be {\it selective} to a type of
molecule.

The process of particle adsorption from the surface into the pit
is simulated on using molecular dynamic technique on the example
of argon atom and graphite surface. The probability of adsorption
from the surface into the pit is calculated with the help of
this simulation. The calculated value of this probability is
used to estimate both efficacy parameters $\gamma$ and $\Gamma$
of surface modification for the case of argon atom absorbed
on graphite surface with a pit.

Some possibilities for nanotechnological applications of
adsorption on the surface with artificial defects are discussed
in detail: fabrication of sensors for trace molecule detection,
separation of isomers, and pattern transfer as a result of
particle self-organization on the modified surface.

\section{Theory of adsorption on a surface with defects.}

Let us consider the system containing the particles of one type
in three states: in a gas, in adsorbed states on defects
(adsorption site D) and on areas of the surface which are free
of defects (adsorption site S). We restrict our consideration to
the case of nanolocal defects with only one adsorption site. The
process of an adsorption of a particle on a surface can be
considered as a chemical reaction

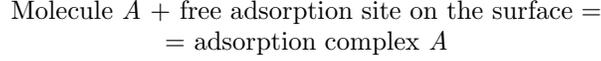
\begin{center}
     Molecule {\it A} + free adsorption site on the surface =

                     = adsorption complex {\it A}
\end{center}

The condition for equilibrium of this "chemical reaction" is

\begin{equation}
\label{2}       \mu_g +\mu_a =\mu'_a
\end{equation}

\noindent where $\mu_g$, $\mu_a$ and $\mu'_a$ are chemical
potentials of the particle in the gas, of the adsorbent per one
adsorption site and of the adsorption complex, respectively. We
shall restrict our theoretical consideration to the case of
physical adsorption. In this case the following approximation
can be used \cite{21}

\begin{equation}
\label{3}                   \mu'_a=\mu_a +\mu_s
\end{equation}

\noindent where $\mu_s$ is the chemical potential of the
particle adsorbed on site S. We use the analogous approximation
for particle adsorbed on the defect. Taking into consideration
this approximations, the condition for thermodynamic equilibrium
of the system is

\begin{equation}
\label{4}                 \mu_g =\mu_d  =\mu_s
\end{equation}

\noindent where $\mu_d$ is the chemical potential of the
particle adsorbed on the defect. This case is analogous to
adsorption of $N_1$ immobile particles adsorbed on $N_2$
adsorption sites. Therefore the chemical potential of the
particle adsorbed on the defect is (see, for example, \cite{21})

\begin{equation}
\label{5}                \mu_d = kT \ln \frac{\nu_d}{(1-\nu_d) q_d}
\end{equation}

\noindent where $\nu_d = N_1 / N_2$ is the filling of adsorption sites
on defects, $q_d$ is the partition function of the particle
adsorbed on the defect, $q_d$ includes internal degrees of freedom,
vibrations, and rotations or frustrated rotations relative to the
surface, and interaction with the surface. The chemical potential
of the particle in the ideal gas is

\begin{equation}
\label{6}   \mu_g = -kT \ln \left( q^{-1}_{int_g} \frac{P}{(kT)^{5/2}}
\left( \frac{h^2}{2\pi m} \right)^{3/2} \right)
\end{equation}

\noindent where $q_{int_g}$ is the partition function of the
particle in the gas, $q_{int_g}$ includes internal degrees of freedom and
rotations, $m$ is the mass of particle, $P$ is the pressure of
particles in the gas. In the case of ideal gas the Langmuir
isotherm for concentration $n_1$ of particles adsorbed on the
defects follows from the equation $\mu_g = \mu_d$

\begin{equation}
\label{7}   n_1 = \nu_d n_d = \frac {q_d \exp (\mu_0/kT)n_d P}
{1 + q_d \exp (\mu_0/kT)P}
\end{equation}

\noindent where $n_d$ is the concentration of defects, and

\begin{equation}
                 \mu_0 = - kT \ln \left( q^{-1}_{int_g} \left( \frac
{2\pi mkT} {h^2} \right)^{3/2} kT \right)
\end{equation}

For some nanotechnological applications it is necessary to increase
the concentration of certain particles adsorbed on a surface.
Some of these applications, namely, fabrication of sensors for
trace molecule detection and separation of isomers will be
discussed below. We believe that following modification of
surface may be useful for such applications: On the surface the
nanolocal artificial defects are produced so that adsorption of
certain particles on defects (sites D) causes significant
increase in the concentration of these particles in comparison
with concentration of these particles on a perfect surface (where
only sites S are present). Such modification is worthwhile only
in the case of low concentration of this particles on perfect
surface, i.e. when filling $\nu_0$ of sites S is small ($\nu_0
\ll 1$). By this is meant that the contribution of particles
adsorbed on sites S into the total concentration of particles
adsorbed on a modified surface is negligible. Total concentration
$n_t$ on modified surface is approximately equal to concentration $n_1$
of particles adsorbed on sites D. Let us introduce the
surface modification efficacy parameter $\gamma$ equal to the
ratio between concentrations on modified surface $n_t$ and
perfect surface $n_2$

\begin{equation}
\label{8}          \gamma  = \frac {n_t} {n_2} = \frac{\nu_d n_d}
{\nu_0 n_0}
\end{equation}

\noindent where $n_0$ is the concentration of sites S on the
perfect surface. It is worthwhile to modify the surface if
$\gamma \gg 1$. Let us obtain the expression for the surface
modification efficacy parameter $\gamma$ for different cases.
The magnitude of this parameter is defined by the equation
$\mu_d = \mu_s$ and therefore it does not explicitly depend on
the gas pressure.

Let us consider the case of low temperature $kT \ll \Delta E$, $\Delta
E$ being the height of the potential barrier between two
adjacent adsorption sites S. In this case particles adsorbed on
sites S jump between these sites. The expression for the
chemical potential of the particle adsorbed on sites S has the
same form as for the particle adsorbed on the defect. In result
the equation $\mu_d = \mu_s$ leads to the relation

\begin{equation}
\label{9}       \frac {\nu_d} {(1-\nu_d) q_d} = \frac {\nu_0} {(1-\nu_0) q_0}
\end{equation}

\noindent where $q_0$ is the partition function of the particle
adsorbed on site S. Now we substitute in Eq. (\ref{9}) the
equalities $\nu_d = n_1/n_d$ and $\nu_0 = n_2/n_0$ and consider
the case $\nu_0 \ll 1$, when the surface modification is
worthwhile for the purpose of selective adsorption. In result we
get

\begin{equation}
\label{10}       \frac {n_1} {n_d -n_1} = \frac {n_2 q_d} {n_0 q_0}
\end{equation}

From Eq. (\ref{10}) we obtain

\begin{equation}
\label{11}         \gamma = \frac {n_d q_d} {n_0 q_0} \left( 1+
\frac {n_2 q_d} {n_0 q_0} \right)^{-1}
\end{equation}

The partition functions of the particles adsorbed on sites D and S
respectively may be represented by the expressions \cite{21}:

\begin{equation}
\label{12}       q_d = q'_d \exp (E_d/kT),
\end{equation}

\begin{equation}
\label{13}       q_0 = q'_0 \exp (E_0/kT),
\end{equation}

\noindent where $q'_d$ and $q'_0$ are the partition functions of the
adsorbed particles including internal degrees of freedom,
vibrations, and rotations or frustrated rotations; $E_d$ and
$E_0$ are adsorption energies of particles adsorbed on sites D and S
correspondingly. We get

\begin{equation}
\label{14}      \gamma = \frac {\beta n_d} {1 + \beta n_2},
\end{equation}

\begin{equation}
                \beta = \frac {q'_d} {n_0 q'_0} \exp \left( \frac
{E_d - E_0} {kT} \right)
\end{equation}

Now we consider the case of high temperature $kT \ll \Delta E$
when particles adsorbed on sites S freely move over the surface.
In the case of low concentration of particles adsorbed on perfect
surface ($\nu_0 \ll 1$) the interaction between adsorbed
particles can be disregarded. Therefore we consider the system of
particles adsorbed on perfect surface as the two-dimensional ideal
gas. Then the chemical potential of the particle adsorbed on
sites S is given by

\begin{equation}
\label{15}        \mu_{s}   = kT \ln \left( \frac{n_2 h^2}
{2\pi mkT q'_s \exp (E_0/kT)} \right)
\end{equation}

\noindent where $q'_s$ is the partition function of the adsorbed
particle, $q'_s$ includes internal degrees of freedom, rotations
and one vibration for constrained motion perpendicular to the
surface. From the relation $\mu_d =\mu_{s}$ we have the following
expressions for the parameter $\gamma$

\begin{equation}
\label{16}      \gamma = \frac {\beta n_d} {1 + \beta n_2},
\end{equation}

where the parameter $\beta$ for the case considered is

\begin{equation}
             \beta = \frac {q'_d h^2} {2\pi mkT q'_s}
\exp \left( \frac {E_d - E_0} {kT} \right)
\end{equation}

Note that in both cases, when adsorbed particles are localized
and freely move over the perfect surface, the dependence of the
parameter $\gamma$ on the concentrations $n_2$ and $n_d$ has
the analogous form (Eqs. (\ref{14}) and (\ref{16})).

To estimate the value of parameter $\gamma$ with the help of Eqs.
(\ref{14}) and (\ref{16}) it is necessary to calculate the
partition functions of the particles. Nevertheless equivalent
results can be obtained using kinetics principles. Contrary to
our derivation of expression for parameter $\gamma$ with the help
of thermodynamical approach, where we use approximation (\ref{9})
valid only for physical adsorption, the consideration of kinetics
below is valid for both physical and chemical adsorption. By
analogy with the standard premises of the Langmuir and BTE
adsorption theories \cite{12}, we suppose that the kinetic
coefficients for the adsorption and desorption of the particles
are independent of their concentration. We suppose also that each
site D on a modified surface is surrounded by area with sites S
where local concentration of particles on this area equal to
their concentration on perfect surface $n_2$. According to
detailed balance for particles exchange between D and S sites we
have

\begin{equation}
\label{17}       k^-_d (1 - \nu_d) n_2 = k^+_d \nu_d n_d
\end{equation}

\noindent where $k^-_d$ is a coefficient of the adsorption on
site D of the particle which have been just adsorbed on site S,
and $k^+_d$ is a coefficient of desorption from site D with
subsequent adsorption on site S. From Eq. (\ref{17}) we obtain
the expression for parameter $\gamma$

\begin{equation}
\label{18}       \gamma = \frac {n_d k^-_d} {n_2 k^+_d} \left(1 +
\frac {k^-_d} {k^+_d} \right)
\end{equation}

The coefficient $k^+_d$ is given by the Arrhenius formula \cite{22}:

\begin{equation}
\label{19}       k^+_d = \Omega_d \exp (-\Delta E_d / kT),
\end{equation}

\noindent where $\Omega_d$ is a frequency multiplier equal in
order of magnitude to the vibration frequency of the particle
adsorbed on the defect, and $\Delta E_d = E_d - E_0 + \Delta E_1$
is the activation energy for the desorption of the particle from
the defect onto the surface free of them, $E_d$ and $E_0$ being
the adsorption energy of the particle adsorbed on sites D and S,
correspondingly, and $\Delta E_1$ is the height of the potential
barrier between adsorption sites on the defect and on the
surface.

Where $kT \gg \Delta E$, the particle freely move over the
surface, and the coefficient of adsorption on a defect site D
from an area of surface with sites S is

\begin{equation}
\label{20}       k^-_d = 2k'R<v>_s n_2 \exp(-\Delta E_1/kT),
\end{equation}

\noindent where $k'$ is the probability of the adsorption on the
defect site D for the particle moving over the surface just after its
collision with site D, $R$ is the radius of the defect, and
$<v>$ the average velocity component parallel to the surface for
the particles adsorbed thereon. We define the probability $k'$
as the ratio $k' = N_a/N_c$ between number $N_a$ of events of
adsorption on site D immediately after the collision and number $N_c$
of these collisions. This quantity is analogous to the
coefficient of attachment of the particles to the surface in the
case of an adsorption from a gas on a surface. After substitution
of Eqs. (\ref{19}) and (\ref{20}) into Eq. (\ref{18}) one have

\begin{equation}
\label{21}           \gamma = \frac {\beta n_d} {1 + \beta n_2},
            \beta = \frac {k'R<v>_s} {\Omega_d}
\exp \left( \frac {E_d - E_0} {kT} \right)
\end{equation}

Where $kT \ll \Delta E$, the particles jump between adjacent
adsorption sites, and in this case $k^-_d = k'l_n \omega \nu_0$,
where $k'$ is the probability of the absorption on the defect
immediately after a jump on this site, $l_n$ is the number of
the adsorption sites S adjacent to the defect,
$\omega$ is the jump frequency defined by the
Arrhenius formula

\begin{equation}
\label{22}       \omega = \Omega_0 \exp \left( - \frac {\Delta E_1}
{kT} \right),
\end{equation}

\noindent where $\Omega_0$ is a frequency multiplier equal in order of
magnitude to the vibration frequency of the particles adsorbed on
the site S. Note that according to simulation \cite{23} the
majority of jumps are between adjacent
adsorption sites. In this case

\begin{equation}
\label{23}           \gamma = \frac {\beta n_d} {1 + \beta n_2},
            \beta = \frac {k'l_n \Omega_0} {n_0 \Omega_d}
\exp \left( \frac {E_d - E_0} {kT} \right)
\end{equation}

Expressions (\ref{14}), (\ref{16}), (\ref{21}), (\ref{23}) can
be extended to the case of some different kinds of defects.
Disregarding the contribution of particles adsorbed on sites
S to the total concentration $n_t$ on of particles adsorbed
modified surface we get

\begin{equation}
             n_t = \sum_i n_{1i}
\end{equation}

\noindent $n_{1i}$ is the concentration of particles adsorbed on
the $i$-th kind of defects. In this case the surface
modification efficacy parameter $\gamma$ for the total surface
is

\begin{equation}
             \gamma = n_t/n_2= \sum_i n_{1i}/n_2 = \sum_i \gamma_i
\end{equation}

\noindent where $\gamma_i$ is the surface modification efficacy
parameter for the $i$-th kind of defects.

For some nanotechnological applications a high filling $\nu_d
\approx 1$ of sites D and absence of adsorbed particles on sites
S are necessary. For example, let us consider a pattern drown by
the particles adsorbed on sites D. To distinguish such pattern
two conditions should be fulfilled. Firstly, a high filling
$\nu_d \approx 1$ of sites D corresponds to a large value of
parameter $\alpha_1 = 1/1-\nu_d$. Secondly, for better pattern
contrast the filling $\nu_d$ of sites D should be significantly
greater than the filling $\nu_0$ of sites S, that corresponds to
a large value of parameter $\alpha_2 = \nu_d/\nu_0$. Let us
introduce the second surface modification efficacy parameter
$\Gamma$ which is equal to product of parameters $\alpha_1$ and
$\alpha_2$

\begin{equation}
\label{24}    \Gamma = \alpha_1 \alpha_2 = \frac {\nu_d} {\nu_0 (1 - \nu_d)}
\end{equation}

The analytical expressions for parameter $\Gamma$ are derived
with the help of both thermodynamical and kinetic approaches.
Under the thermodynamical consideration we use the equation
$\mu_d = \mu_s$ analogous to the derivation of expressions for
parameter $\gamma$. In the case of jump diffusion of particles
adsorbed on sites S we get

\begin{equation}
\label{25}    \Gamma = \frac {q'_d} {q'_0} \exp \frac {E_d - E_0} {kT}
\end{equation}

When the particles adsorbed on sites S freely move
over the surface we get the following expression for parameter
$\Gamma$

\begin{equation}
\label{26}     \Gamma =
\frac{n_0 h^2 q'_d} {2\pi mkT q'_0} \exp \frac {E_d-E_0} {kT}
\end{equation}

The Eqs. (\ref{25}) and (\ref{26}) are not convenient for
practical use (analogous to Eqs. (\ref{14}) and (\ref{16}))
because it is necessary to calculate the partition functions.
Therefore we have derived the equivalent expressions with the
help of kinetic consideration. In the case of jump diffusion of
particles adsorbed on sites S we get

\begin{equation}
\label{27}     \Gamma = \frac {k'l_n \Omega_0} {\Omega_d} \exp
\frac {E_d - E_0} {kT}
\end{equation}

In the case when particles adsorbed on sites S freely move
over the surface we obtain the following expression

\begin{equation}
\label{28}     \Gamma = \frac {k'R<v>_s n_0} {\Omega_d} \exp
\frac {E_d - E_0} {kT}
\end{equation}

Note, that nanotechnological procedures described by
parameters $\gamma$ and $\Gamma$ are applied only to the case of
low concentration of adsorbed particles on sites S
($\nu_0 \ll 1$). Therefore the neglect of interparticle
interaction is adequate for our consideration. However, Eqs.
(\ref{21}), (\ref{23}), (\ref{27}) and (\ref{28}) derived in the
framework of the kinetic approach may be easily extended to the
case where this interaction is important (analogous to
homogeneous adsorption, see, for example, \cite{12aa}). To take
into account an interparticle interaction it is sufficient to
replace $E_d$ by $E_d+w_dN_d$ and $E_0$ by
$E_0+w_0N_0$, where $w_d$ and $w_0$ are interaction energies
between neighbor particles, and $N_d$ and $N_0$ are average numbers
of nearest neighbors for particles adsorbed on sites D and S,
respectively.

\section{Adsorption Energy of molecules in the pits.}

The adsorption of particles on an unperfect surface for the
simplest cases had been studied previously: in the cylindrical
pore \cite{16}, in the split-shaped pore \cite{17}, at the
intersection of two perpendicular steps forming a reentrant
corner \cite{17}. Nevertheless these studies dealt only
with rather large surface defects. The case of nanolocal
defects with sizes of order of the size of only one adsorption
site was not considered.

Theory presented above does not depend on the nature of defects.
They may be structural defects or chemical one and so on. We
consider molecule adsorption on a single type of structural defects,
namely, pits made by removing tens of atoms from the surface.
Our interest in this type of defects is attracted by following
reasons. Firstly, according to expressions (\ref{14}), (\ref{16}),
(\ref{21}), (\ref{23}) surface modification efficacy parameter
$\gamma \sim \exp ((E_p - E_0)/kT)$. Therefore the modification of
surface by presence of the pits with certain size can cause the
increase of concentration of certain kind of molecules. Some
nanotechnological applications based on this effect are discussed
in Sec. 4 (fabrication of sensors for trace molecule detection,
separation of isomers {\it etc.}). Secondly, because of the recent
nanotechnological advances the various techniques to produce pits
on the surface had been elaborated. For example, it is possible
to fabricate the pits only several nanometers in diameter
\cite{1b,1c} and even to remove a single atom from the surface
structure \cite{1c} with the help of field desorption by STM tip.
Therefore a theoretical study of particle adsorption in a pit is
an actual problem now.

Here we have calculated the adsorption energy $E_p$ for the argon
atom and benzene and acetylene molecules adsorbed on graphite
surface in pits that are different in size but can accommodate
only one particle. We have choose a graphite for our study
because the adsorption of argon (see, for example, \cite{18}) and
various molecules (see, for example, \cite{18a}--\cite{18c}) on
perfect graphite surface have been carefully theoretically
studied in a set of papers. Besides, the pits on the graphite
surface were produced by nanolocal chemical reactions near STM
tip \cite{2b} and by bombardment of cations with subsequent
etching \cite{12a}.

This is a nonspecific adsorption case where the
adsorbate-adsorbent interaction is of the Van der Waals type
\cite{12}. We have described the interaction between the argon
atom and the carbon atoms in terms of the Lennard-Jones
potential

\begin{equation}
\label{29} U = 4\varepsilon \left[ (\sigma /r)^{12} - (\sigma /r)^6 \right]
\end{equation}

\noindent with a cutoff radius of $r_c=3.2$, where the potential
parameters have been taken to be $\sigma=3.12$ $\AA$ and
$\varepsilon=54.4$ K \cite{18}, $\sigma=3.82 \AA$ and
$\varepsilon=31.6$ K for the C-C interaction \cite{19}, and
$\sigma=3.37$ $\AA$ and $\varepsilon=21.7$ K for the H-C
interaction \cite{19}. The atoms of graphite have been assumed to
be fixed at the lattice sites, three graphite layers with 288
atoms in each layer being taken into account. The interatomic
bonds in the molecules have been taken to be rigid, and the
interatomic bond angles fixed. The bond length values used (1.4
$\AA$ for the C-C bond and 1.08 $\AA$ for the H-C one) have been
borrowed from \cite{19}.

We have considered pits with near circular shape that are created
by the removal of atoms located inside circles of radii $R_1$ and
$R_2$ for first and second graphite layers correspondingly. The
centers or these circles locate on one vertical line $l$. All
possible pits with $0 \le R_2 < R_1 < 5.06 \AA$ for argon atom
and with $0 \le R_2 < R_1 < 6.80 \AA$ for acetylene and benzene
molecules have been investigated. Several different positions of
line $l$ relative atoms of graphite were considered. Namely, the
line $l$ passes through: 1) an atom of upper graphite layer under
which an atom of the second layer locates (type A); 2) an atom of
upper graphite layer under which an atom of the second layer is
absent (type B); 3) the center of a bond between two atoms of
upper graphite layer (type C); 4) the center of hexagon formed by
atoms of upper graphite layer (type D). The optimal pits with one
and two graphite layer deep, where the adsorption energy is a
maximum $E_p^{max}$ are found for all three adsorbates. For all
these cases the energies of adsorption in optimal pits are
maximum when the center of a particle coincide with the center of
the pit. Several pits of oblong shape with the size close to the
size of the optimal pit with near circular shape are also
examined for acetylene molecule. Nevertheless the adsorption
energies in these oblong pits are less than adsorption energy in
the optimal pit with near circular shape. These adsorption
energies $E_p^{max}$ and characteristics of optimal pits are
listed in Table 1 for all three adsorbates.

The calculations show that the adsorption energy in an optimal
pit for small particles (argon atom and acetylene molecule) is
nearly twice as large as the adsorption energy on perfect
surface. The adsorption energy in the optimal pit with two
graphite layer deep is only 7 \% greater than the adsorption
energy in the optimal pit with two graphite layer deep. The
further increase of pit depth does not cause the adsorption
energy to increase. Therefore a pit a mere 1-2 atomic layers deep
may be sufficient to increase the absorption energy to the
utmost, which is fairly convenient for nanotechnological
applications.

A particle adsorbed in a pit interacts with walls and bottom of
the pit. And a particle adsorbed in a capillary interacts only
with walls of capillary. Therefore the adsorption energy of a
particle in a pit is greater than in a capillary with the same
radius. Previously the adsorption energy of a particle in a
cylindrical capillaries with various radii has been calculated
analytically \cite{16}. In \cite{16} the interaction between the
particle and the walls of the capillary has also been described
in terms of the Lennard-Jones potential. However, the capillary
walls in this simplified model have been considered continuous.
According to calculations in the framework of this model the
ratio $\epsilon_0$ between the maximal adsorption energy
$E_c^{max}$ of a particle in the capillary and the adsorption
energy on the perfect surface is $\epsilon_0 =
E_c^{max}/E_s\simeq3$. This value of $\epsilon_0$ is 1.5 times as
great as value calculated by us for adsorption in a pit
$\epsilon_0 = 2.06$. Therefore the consideration of the location
of all the atoms of surface is necessary for exact calculation of
adsorption energy in a pit or a pore. For benzene molecule
adsorbed in an optimal pit, the simplified model \cite{16}
overestimates the maximal increase in the adsorption energy even
more because atoms of adsorbed molecule interact weaker with pit
wall sections distant from them than with the pit bottom. The
value of the energy ratio $\epsilon_0$ calculated here for argon
atom adsorption in a pit on graphite surface is close to the one
calculated for argon atom adsorption on microporous magnesium
oxide with taking into account location of all the atoms
\cite{17}. According to the calculation \cite{17} $E_p^{max}/E_s$
$\simeq 1.8$ for the split-shaped pore and at the intersection of
two perpendicular steps forming a reentrant corner.

Fig. 1 demonstrates that the energy of adsorption in a pit is
determined by the size of the pit. This figure also shows that a
pit too small to accommodate a molecule even reduces its
adsorption energy. This two facts open up new interesting
possibilities for nanotechnological applications which have been
discussed below. An additional point to emphasize is
that the "flat" benzene molecule placed in a pit of optimal size
but with a "unflat" bottom has a lower adsorption energy than in
the case of unmodified surface.

\section{Simulation of adsorption of particle in pits.}

The parameters $\gamma$ and $\Gamma$ have been estimated for the
case of argon atom adsorption on graphite surface with pits. For
this estimation it is necessary to calculate the oscillation
frequency $\Omega_p$ of atom adsorbed in a pit and the
probability $k'$ of the absorption in the pit for the particle
moving over the surface just after its collision with the pit. These
quantities were obtained by molecular
dynamics simulations. According to Eqs. (\ref{21}) and (\ref{28})
parameters $\gamma$ and $\Gamma$ increase with the adsorption
energies $E_p$ in a pit. Therefore two pits were used for
simulation: the optimal pit with maximum adsorption energy in it
and the pit with maximum adsorption energy among the investigated
pits with only one graphite layer depth (see Table 1). The
adsorption energy in this pit is only 8\% less than adsorption
energy in optimal pit.

A system consisting of three graphite layers with 288 atoms in
each layer was used in simulation. Along X and Y axes parallel to
the graphite surface we impose periodic boundary conditions on
the system. The size of the simulation cell in these directions
was $27.08 \times 23.43 \AA$. The interaction between the argon
atom and carbon atoms was represented by Lennard-Jones potential
(\ref{29}). The atoms of second and third graphite layers were
fixed at equilibrium positions. The interaction between carbon
atoms of first graphite layer was described by modified Born
potential

\begin{equation}
U = \frac{ \alpha-\beta}{2} \sum ^{N}_{i,j=1}
( \frac{ ({\bf r}_{ij}-{\bf r}_{0ij}){\bf r}_{ij}}{r_{ij}} )
^{2}+ \frac{ \beta}{2} \sum^{N}_{i,j=1}({\bf r}_{ij}-{\bf r}_{0ij})^{2}
\end{equation}

\noindent where ${\bf r}_{ij}$ are distances between carbon
atoms, ${\bf r}_{0ij}$ are distances between carbon atoms at
equilibrium positions, $\alpha$ and $\beta$ are force constants.
We take $\alpha=505.1$ N/m and $\beta=84.4$ N/m \cite{28}. The
Born potential represents the expansion of the interaction energy
between carbon atoms in terms of $\Delta {\bf r}_{ij} = {\bf
r}_{ij} - {\bf r}_{0ij}$. Therefore this potential is adequate
only for small values of $\Delta {\bf r}_{ij}$. We simulated the
system at temperatures that approximately 20 times less than the
temperature of graphite melting. At such temperatures the
displacements of graphite atoms from equilibrium positions and,
consequently, values of $\Delta {\bf r}_{ij}$ are small.
Therefore we consider that Born potential is acceptable for our
simulation.

The equations of motion were integrated using the leap frog
algorithm. The integration step used was $\tau=2.\cdot 10^{-15}$
s. Initially the system of graphite atoms has been brought to the
equilibrium during $5\cdot10^3$ steps (about 300 oscillations of
atoms in graphite) in canonical ensemble and $5\cdot10^3$ steps
in microcanonical ensemble. Further simulations were performed in
microcanonical ensemble. The total energy of the system with the
temperature 185 K was conserved to within 1 \% and average
fluctuations of temperature were within 3 \%.

The estimation of oscillation frequencies $\Omega_d$ of atom
adsorbed in the pits has been performed for the system with the
temperature 165 K. Initially we took the system without a pit
with equilibrium coordinates and velocities of carbon atoms. Then
the pit arises near the center of simulation cell and an atom was
placed in the pit at the position that corresponds to the maximum
adsorption energy. The new system with the pit and the atom
adsorbed in this pit has been brought to the equilibrium during
$2\cdot10^3$ steps. Then frequencies $\Omega_d$ were calculated
during $6\cdot10^3$ steps. The obtained values averaged over 40
modeling experiments are $\Omega_{d1}=5.3\cdot10^{11}$ s$^{-1}$
and $\Omega_{d}=4.7\cdot10^{11}$ s$^{-1}$ for optimal pits with
one layer and two layers deep.

The size of simulation cell used is not sufficient for argon atom appeared
in any place of the cell to come to equilibrium with the surface
before a collision with the pit. To overcome this difficulty the
simulation of the process of the atom adsorption from the surface
into the pit was carried out using the following
procedure. An argon atom is placed in the center of simulation
cell without a pit at the position that corresponds to the
minimum of interaction energy with the surface. Then this system
comes to equilibrium until the argon atom reaches the boundary of
the cell. At the instant the argon atom crosses the boundary the
pit arises in the center of the cell. The appearance of the pit
was investigated for the system at the temperature 185 K. The
change in the energy of interaction between argon atom and
graphite atoms caused by the appearance of the pit is within
0.03\%. Therefore we consider that the appearance of the pit does
not essentially perturb the motion of argon atom and equilibrium
between argon atom and surface. During the time between the
appearance of the pit and the argon atom collision with the pit
graphite atoms make about 100 oscillations that is sufficient for
system to come to equilibrium \cite{7}.

On the average an argon atom makes 12 jumps before collision with
the pit. Average change in the kinetic energy of argon atom
in result of collision with the surface ~is $<\Delta E>=17\pm1.2$\%,
$\Delta E_i =2|E_i-E_{i-1}|\cdot 100\%/(E_i+E_{i-1})$, where
$E_{i-1}$ and $E_i$ are average through the time of jump kinetic
energies of argon atom between $i$-1-th and $i$-th, and $i$-th and
$i$+1-th between collisions with graphite surface respectively.
The value $<\Delta E>$ has been obtained by averaging over 20 argon atoms,
each making 10 jumps along the surface. We believe that observed
energy exchange between argon atom and the surface is sufficient
to argon atom to come to equilibrium with the surface.

An argon atom was considered to collide with the pit when the
distance between it and the center of the pit was less than the
distance between the center of the pit and a adjacent adsorption
site S. An argon atom was considered to be adsorbed in the pit
when it makes two oscillations in the pit. The simulation with one
atom was performed until the atom reflects from the pit, adsorbs
into the pit or evaporates from the surface. To exclude atoms
with low velocities the time of experiment was limited to
$2\cdot10^4$ integration steps. The results of simulations are
averaged over all modelling experiments for given temperature of
surface. These results and calculation of the probability $k'$ of
the adsorption into the pit and the efficacy parameters are
presented in Table 2. The probability $k'$ of the adsorption into
the pit slightly increases with temperature. We offer the
following explanation for this increase. When a particle occurs
on the surface near a pit edge it interacts with a less number of
atoms of the surface than a particle located on perfect surface.
Thus the pit is surrounded by an energetical barrier. The
magnitude of this barrier is found to be about 300 K in simulated
system. Therefore the fraction of particles capable to overcome
the barrier increases with the increase of temperature for the
considered temperature range 120-280 K.

According to our estimations for considered case of argon atom
and optimal pit in a graphite surface at temperature 200 K we get
$\beta n_0 \ll 1$ for $\nu_0 \ll 1$. Then Eq. (\ref{21}) is simplified

\begin{equation}
\label{30}        \gamma=\beta n_d
\end{equation}

\noindent so that the parameter $\gamma$ does not depend on the
concentration of atoms in the gas. The Eq. (\ref{30}) have been
used to estimate the parameter $\gamma$ for considered case. Let
us express the concentration of pits in terms of fraction of the
surface occupied by them, i.e., $n_d = \alpha/\pi R^2$. The
parameter $\gamma$ is calculated for $\alpha = 0.1$. The
calculated parameters $\gamma$ and $\Gamma$ are presented in Table 2.
The magnitudes of these parameters are sufficient for
nanotechnological applications discussed below. We believe that
this conclusion may be also obtained for various other pairs
adsorbent-adsorbate.

Graphite is a very homogeneous adsorbent, i.e., the potential
barriers $\Delta E$ between the adjacent adsorption sites on it
are low. For the argon atom, $\Delta E / k \approx 32$ K \cite{24}.
At low temperatures ($kT < \Delta E$) the submonolayer ($\nu_0 \ll 1$)
adsorption of argon on a graphite surface is impossible for any
experimentally obtainable concentrations of atoms in the gas.
Therefore the case of jump diffusion of adsorbed particles is not
considered for this pair adsorbent-adsorbate.

\section{Nanotechnology applications.}

Here we describe some possible nanotechnology applications for
the adsorption on a surface with nanometer-scale artificial
defects (see also \cite{24aa}): pattern transfer as a result of
self-organization of particles deposited on such surface, and
using of selective adsorption for sensors fabrication and isomer
separation.

{\bf Pattern transfer.} 'he elaboration of methods of pattern
transfer in nanometer scale, i.e. fabrication of nanostructures
on the surface according to a given scheme is very important for
some nanotechnology applications (for example, superdense
recording of information, fabrication of periodic arrays of quantum dots and
quantum wires, one-electron devices {\it etc.}).
In principle, surface nanostructures had been produced with the
help of STM or AFM tip by the transfer of single atoms \cite{29,29a},
clusters \cite{30,30a}, and nanoparticles \cite{31}. Nevertheless,
until now these methods have the productivity insufficient for
nanotechnology purposes.

The other set of methods for surface nanostructures fabrication
is based on self-organization of deposited particles: a) the
formation of islands in the result of nucleation in two-dimensional
film (see, for example, \cite{4}); b)self-organizing ordering in
epitaxial layer with facets formation (see, for example,
\cite{34}); c) the formation of periodical surface structure by
laser radiation \cite{35,36}. The method a) allows to control
only average distance between islands. The methods b) and c) are
suitable only for periodical nanostructures fabrication with periods
by chemical composition of nanoobjects and wave length of laser
radiation, correspondingly. The productivity of these methods is
considerably greater than transfer of single particles by STM or
AFM tip. Nevertheless they are not suitable for pattern transfer.

A particle adsorbed on defect may play the role of nucleation
center for island formation from the particles deposited on the
surface. For example, the gold islands on the graphite surface
irradiated with ions form only around defects \cite{4}. We propose
to deposit particles on the surface with defects located with the
help of STM or AFM according to given scheme. In this case
the formed islands would merge into planned surface nanostructure.

Recently the diode on one molecule was fabricated \cite{37}.
We proposed to place such molecules in electronic schemes
of nanometer scale with the help of adsorption on
specially created defects. We believe that by these means
it is possible to control not only the position but
also the orientation of the molecule.

{\bf Sensors for molecular detection.}
Methods for detecting traces of certain molecules in a gas where
their concentration is low are of great importance in modern
technology and find application, for example, to ensure safety
in chemical industries and monitor environmental pollution. The
further development of such methods, particularly is therefore a
high-priority task. A weak spot in methods used is the collection
of molecules from air. The known method of improving sensitivity
\cite{10} involves the nonselective accumulation of the trace
molecules of interest on a cooled substrate and their subsequent
pulsed laser desorption. It therefore seems extremely tempting to
develop a method for {\it selective} adsorption of molecules on a
cold surface.

Calculations performed show that the adsorption energy in a
pit is determined by the size of the pit. We propose to modify
the surface of sensor by presence of the pits with maximum
adsorption energy for certain kind of molecules.
According to the theory presented the concentrations of these
molecules on the surface of sensor will increase. Therefore it
would enable one to add one more selective step to the existing
two detection steps --- the optical and the mass-spectrometric
ones.

{\bf Isomer separation.} Pits can be arranged on a
surface so closely that the distances between them have the same
order as their size. The molecular adsorption energy on such a
surface will be higher for molecules that fit in the pits,
compared to that on the unmodified surface, and lower for those
which fail to fit in the pits. A surface modified in this
fashion could be used to effect the selective adsorption of a
particular molecular species from a mixture of different
molecules. What is more, a surface can be modified to have pits
capable of accommodating only one of several isomeric molecular
species. We suggest using such a modified surface to detect or
separate molecules differing in {\it shape} only (e.g., to
isolate linear or cyclic isomers from their mixture and separate
fullerene isomers and left- and right-handed molecules).

In summary, the theory of adsorption on the surface with
nanolocal defects is developed. Two efficacy parameters of
surface modification by creation of artificial defects are
introduced for different nanotechnological applications.
The estimations with the help of molecular dynamics
simulations on the example of argon atom adsorption in the pit
on the graphite surface show that magnitudes of these parameters
are sufficient for possible nanotechnological applications.

\section*{Acknowledgement.}

We are grateful to Referee for useful comments improving the text.
This work was supported by the Program "Surface atomic
structures", and by the grant of RFBR. The work of Popov A.M. has
been made possible by a fellowship of International Center for
Fundamental Physics in Moscow.

\newpage
\begin{center}
{\bf Figure captions.}
\end{center}

Fig.1. The dependence of the energy $E_p$ of adsorption in a pit
with one graphite layer deep on the number $N_1$ of atoms removed
when making this pit; a) acetylene molecule, b) benzene molecule.
Solid squares corresponds to pits of type A and open circles
correspond to pits of type C. Points corresponds to pits of
types B and D are not shown because they couincide with
points corresponds to pits of types A and C.

\newpage
{\bf Table 1.} Characteristics of optimal pits: the type of the
pit $l$ (see text); the number of atoms removed when making this pit from
the first and second graphite layers, $N_1$ and $N_2$,
respectively. Adsorption energies in the optimal pit $E_p^{max}$
and in the optimal pit relative to that on the unmodified
surface $\Delta E = E_p^{max}-E_0$ (in Kelvin degrees); the energy ratios
$\epsilon_0 = E_p^{max}/E_0$ and $\epsilon_1 = E_p^{max}/E_{p1}^{max}$,
where $E_{p1}^{max}$ is the maximum adsorption energy among the
pits with one graphite layer deep.

\bigskip

\bigskip

\begin{tabular}{|l|c|c|c|c|c|c|c|}\hline
  &     &     &      &      &      &      &    \\
particle & $l$ & $N_1$ & $N_2$ & $E_p^{max}$ & $\Delta E$ & $\epsilon_0$ &
$\epsilon_1$ \\
  &     &     &      &      &      &      & \\ \hline
  &     &     &      &      &      &      & \\
argon atom & A & 13 & 12 & 1751 & 901 & 2.06 & 1.083 \\
  &     &     &      &      &      &      & \\ \hline
  &     &     &      &      &      &      & \\
acetylene molecule & C & 16 & 16 & 3682 & 1873 & 2.03 & 1.108 \\
  &      &      &      &      &      &      & \\ \hline
  &     &     &      &      &      &      & \\
benzene molecule & B & 31 & 25 & 8379 & 2944 & 1.54 & 1.069 \\
  &      &      &      &      &      &      & \\ \hline
\end{tabular}

\bigskip

\bigskip

{\bf Table 2.} \quad Results of simulation of adsorption of atom
from the surface into the pit and calculation of surface
modification parameters. $h$ is the depth of pit in graphite
layers; $T_g$ is the temperature of graphite surface; $N_c$ is
the number of atoms collided with the pit; $N_a$ is the number of
atoms adsorbed into the pit; $k'$ is the probability of the
absorption into the pit, $\gamma$ and $\Gamma$ are surface
modification parameters. The accuracy of calculation of $k'$ and
parameters $\gamma$ and $\Gamma$ is connected with statistical error
of detected events of atom adsorption into the pit.

\bigskip

\bigskip

\bigskip

\begin{tabular}{|c|c|c|c|c|c|c|}\hline
   &     &     &    &                 &                 &  \\
$h$ & $T_g$ & $N_c$ & $N_a$ & $k'$ & $\gamma$           & $\Gamma$ \\
   &     &     &    &                 &                 &  \\ \hline
   &     &     &    &                 &                 &  \\
 1 & 123 & 502 & 62 & $0.124\pm0.017$ & $24.16\pm 3.25$ & $3438.0\pm 462.8$\\
 1 & 165 & 350 & 51 & $0.146\pm0.022$ & $ 6.90\pm 1.03$ & $ 981.5\pm 147.1$\\
 1 & 205 & 602 & 93 & $0.154\pm0.017$ & $ 3.28\pm 0.37$ & $ 467.1\pm  52.0$\\
 1 & 248 & 631 & 93 & $0.147\pm0.016$ & $ 1.81\pm 0.20$ & $ 258.2\pm  28.7$\\
 1 & 286 & 662 & 112& $0.169\pm0.017$ & $ 1.49\pm 0.15$ & $ 211.8\pm  21.6$\\
 2 & 123 & 367 & 41 & $0.112\pm0.018$ & $72.72\pm11.97$ &$10347.9\pm1704.0$\\
 2 & 165 & 408 & 40 & $0.098\pm0.016$ & $11.70\pm 1.94$ & $1665.4\pm 275.9$\\
 2 & 205 & 521 & 65 & $0.125\pm0.016$ & $ 5.69\pm 0.75$ & $ 809.5\pm 106.5$\\
 2 & 248 & 593 & 75 & $0.126\pm0.016$ & $ 2.98\pm 0.37$ & $ 424.5\pm  52.0$\\
 2 & 286 & 556 & 101& $0.182\pm0.020$ & $ 2.85\pm 0.31$ & $ 405.5\pm  43.9$\\
   &     &     &    &                 &                &  \\ \hline
\end{tabular}
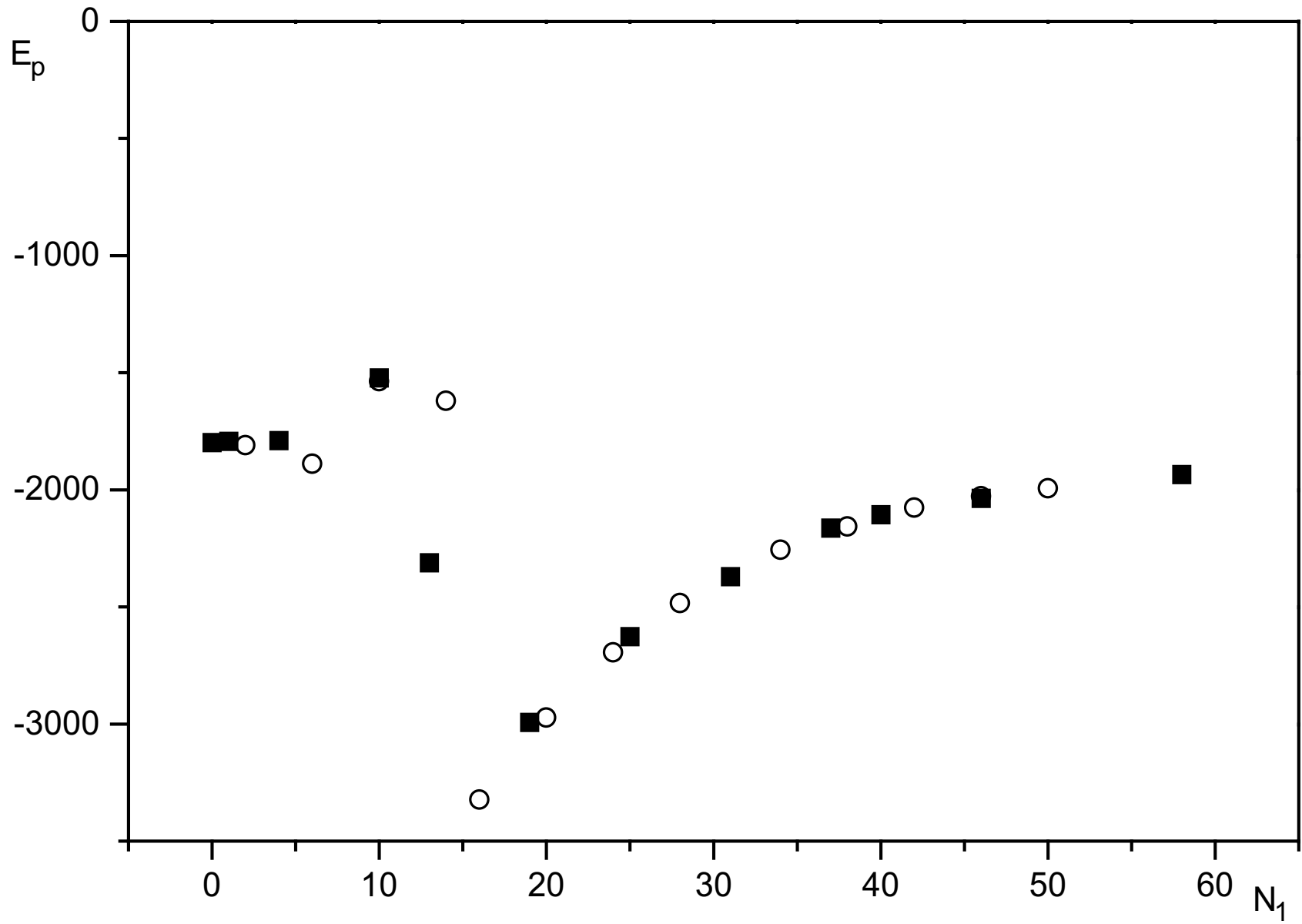
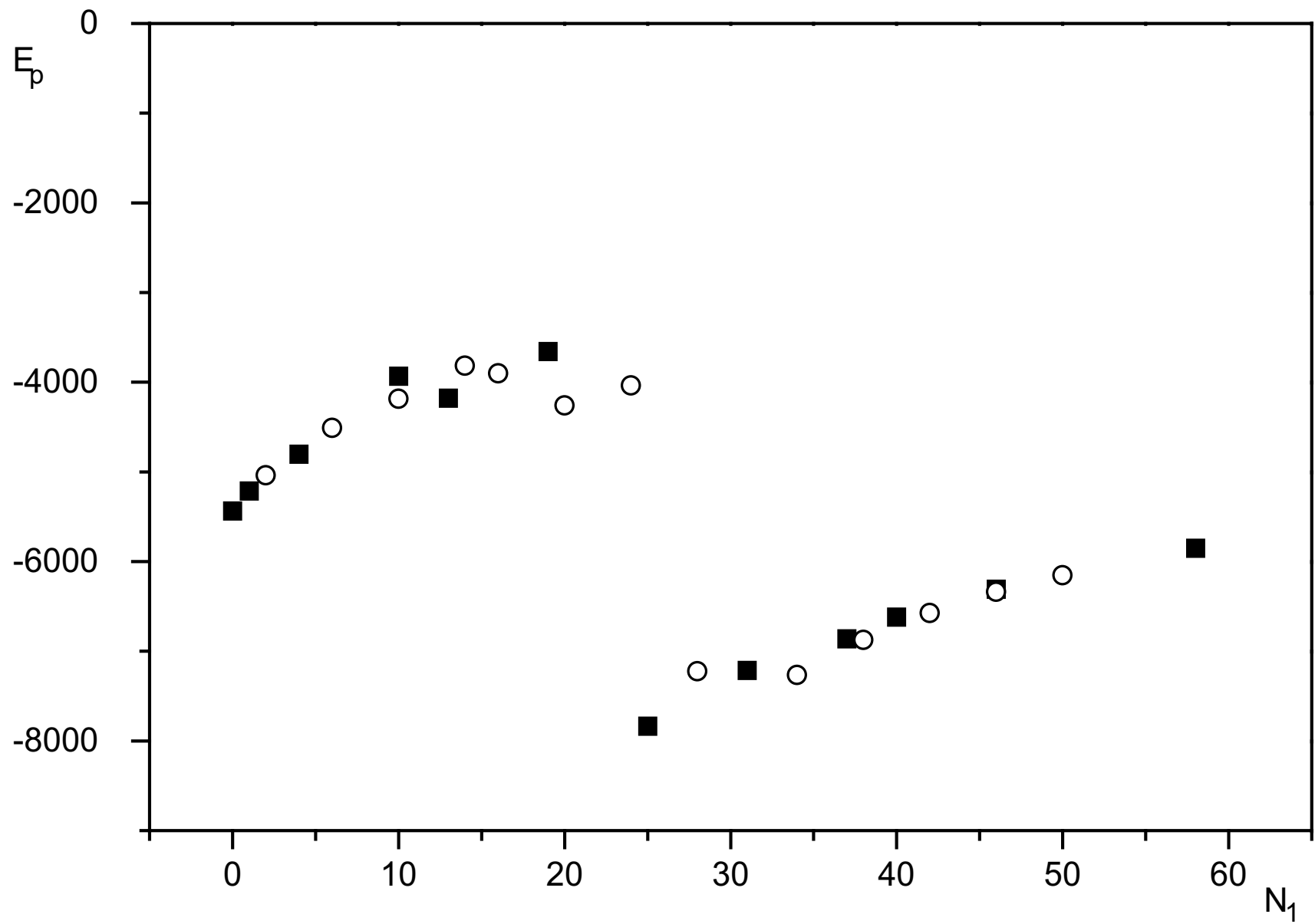
\end{document}